\documentclass[11pt,a4paper]{article}
\pdfoutput=1
\usepackage{jcappub}
\usepackage{bm}

\title{Higher derivative scalar-tensor theory through a non-dynamical scalar field}

\author[a]{Xian Gao,} 
\affiliation[a]{School of Physics and Astronomy,
Sun Yat-sen University, Guangzhou 510275, China}
\emailAdd{gaoxian@mail.sysu.edu.cn}

\author[b]{Masahide Yamaguchi} 
\affiliation[b]{Department of Physics, Tokyo Institute of Technology,
Tokyo 152-8551, Japan} 
\emailAdd{gucci@phys.titech.ac.jp}

\author[c]{and Daisuke Yoshida}
\affiliation[c]{Department of Physics, McGill University, Montr\'eal, QC, H3A 2T8, Canada}
\emailAdd{d.yoshida@physics.mcgill.ca}

\abstract{We propose a new class of higher derivative scalar-tensor theories without the Ostrogradsky's ghost instabilities. The construction of our theory is originally motivated by a scalar field with spacelike gradient, which enables us to fix a gauge in which the scalar field appears to be non-dynamical. We dub such a gauge as the spatial gauge. Though the scalar field loses its dynamics, the spatial gauge fixing breaks the time diffeomorphism invariance and thus excites a scalar mode in the gravity sector. We generalize this idea and construct a general class of scalar-tensor theories through a non-dynamical scalar field, which preserves only spatial covariance. We perform a Hamiltonian analysis and confirm that there are at most three (two tensors and one scalar) dynamical degrees of freedom, which ensures the absence of a degree of freedom due to higher derivatives. Our construction opens a new branch of scalar-tensor theories with higher derivatives.}

\begin{document}
\maketitle

\section{Introduction}

A scalar-tensor theory gives an important framework to describe
inflation and dark energy. The action of a scalar field with the
canonical kinetic term and a potential (together with the
Einstein-Hilbert action) has been long considered as the standard
one. Later, a more general action consisting of an arbitrary function of a
scalar field and its canonical kinetic term was proposed as candidates
to describe inflation dubbed as k-inflation
\cite{ArmendarizPicon:1999rj} and dark energy dubbed as k-essence
\cite{Chiba:1999ka,ArmendarizPicon:2000dh}. This action is the most
general single scalar field one consisting of a scalar field and
its first derivatives.

Then, it is a natural direction to consider an action including higher
derivative terms. Unfortunately, it was shown more than 150 years ago
that a non-singular system with higher-than-first time derivatives
leads to ghost instabilities
\cite{Ostrogradsky:1850fid,Woodard:2006nt,Woodard:2015zca}. Thus, in
order to obtain a healthy theory with higher derivatives, we need to
consider singular (degenerate) system. Recently, the systematic
construction of a healthy point-particle theory with higher time
derivative terms
\cite{Motohashi:2014opa,Motohashi:2016ftl,Klein:2016aiq,Kimura:2017gcy,Motohashi:2017eya,Motohashi:2018pxg}
as well as their field extensions \cite{Crisostomi:2017aim,Kimura:2018sfs}
have been given.

One of the important examples of such a ghost-free field theory is the
Galileon theory in the Minkowski background \cite{Nicolis:2008in}, which
respects the so-called Galilean shift symmetry and gives the most
general action whose (Euler-Lagrange) equations of motion are second
order even though their action has second order derivative
terms. Surprisingly, the scalar-tensor correspondence of this Galileon
theory, which is the most general single field scalar-tensor theory with
second order (Euler-Lagrangian) equations of motion, had been found by
Horndeski more than 40 years ago \cite{Horndeski:1974wa} and this theory
was rediscovered in the context of generalized Galileon theory \cite{Deffayet:2011gz}, of which the equivalence to the Horndeski theory was shown in
\cite{Kobayashi:2011nu}.

However, the Horndeski theory was shown to be not the most general
single field scalar-tensor theory without ghost instabilities, because
the requirement to the Horndeski theory, i.e., the equations of motion are of second order in derivatives, is too strong. In fact, Gleyzes
et al. explicitly constructed theories beyond the Horndeski scope without ghost instabilities \cite{Gleyzes:2014dya,Gleyzes:2014qga}. Langlois and Noui further extended this direction by paying particular attention to the
importance of the degeneracy of the kinetic matrix and proposed the
degenerate higher order scalar tensor (DHOST) theory
\cite{Langlois:2015cwa,Langlois:2015skt,Crisostomi:2016czh,Achour:2016rkg,Crisostomi:2016czh}.
Though the non-trivial branch of DHOST theories suffers from gradient instabilities for cosmological perturbations \cite{deRham:2016wji, Langlois:2017mxy}, the corresponding vector-tensor theory has stable cosmological solutions \cite{Kimura:2016rzw, Kase:2018tsb}.

Another interesting yet more powerful approach to obtaining a healthy scalar-tensor theory with higher derivatives is to focus on specific configurations (vacuum expectation values) of the scalar field, which explicitly breaks the local Lorentz symmetry
(or general covariance). Theories in this direction include the
effective field theory (EFT) of inflation
\cite{Dubovsky:2005xd,Creminelli:2006xe,Cheung:2007st} and
Ho\v{r}ava-Lifshitz theory \cite{Horava:2009uw,Horava:2009if},
where only the spatial symmetries are respected. Interestingly,
when choosing the so-called unitary gauge by fixing $t=\phi(t,\vec{x})$,
the Horndeski Lagrangian can be recast in a form similar to that of the
EFT of inflation for cosmological perturbations
\cite{Gleyzes:2013ooa}.  In \cite{Gao:2014soa,Gao:2014fra}, one of the
authors in this work proposed a general framework to construct
scalar-tensor theories with keeping only spatial symmetries in a
systematic way, which manifestly avoids the ghost instabilities
associated with higher time derivative terms while arbitrarily higher
spatial derivatives can be introduced.\footnote{See
Ref. \cite{DeFelice:2018ewo} for the discussions when this theory
apparently recovers general covariance.}  Recently, further extension
including the time derivative of the lapse function has been made
\cite{Gao:2018znj}. It should be noticed that this theory is constructed
based on the (e.g. constant time) spacelike hypersurface given by a timelike scalar
field. By timelike/spacelike scalar field, we refer to a scalar field
with timelike/spacelike gradient.

In the current work, we will pursue the opposite direction, where the
theory is constructed based on a spacelike scalar field originally. In the next section, we first investigate the Horndeski theory in which the scalar field is spacelike. This
spacelike scalar field allows us to fix a gauge, which we dub as the
spatial gauge, in which the scalar field loses dynamics and appears to be non-dynamical. 
The most important observation is that, once the action of the theory is given,
one can extend our theory by regarding that the scalar field can be not only spacelike but also timelike. As the result, our construction appears to be a novel class of spatially covariant scalar-tensor theories
alternative to those proposed in
\cite{Gao:2014soa,Gao:2014fra,Gao:2018znj}.
The crucial ingredient in our construction is the scalar field, which does
not have a kinetic term from the construction and hence is
non-dynamical. Although this scalar field itself might not have dynamics,
 its presence (together with the gauge fixing) breaks general covariance and thus induces an alternative dynamical
scalar degree of freedom in the gravity sector, which might have novel features.

The idea of having matter
fields with time-independent but space-dependent background
configurations in the cosmological situation
 was firstly introduced in the ``elastic inflation''
\cite{Gruzinov:2004ty} and further systematically investigated in
``solid inflation'' \cite{Endlich:2012pz}, where the scalar fields are
Nambu-Goldstone bosons associated with breaking of spatial
diffeomorphism.  This is similar to the usual EFT of
inflation \cite{Dubovsky:2005xd,Creminelli:2006xe,Cheung:2007st}, where the adiabatic mode
of matter perturbations is associated with the breaking of time
diffeomorphism\footnote{Effective theories with general spacetime symmetry breaking have also been investigated \cite{Leutwyler:1996er,Goon:2012dy,Nicolis:2013lma,Hidaka:2014fra,Nicolis:2015sra,Cannone:2014uqa,Cannone:2015rra,Graef:2015ova,Son:2005ak}.}. 
When apparently recovering the general covariance by employing the St\"{u}ckelberg trick (see e.g., \cite{Germani:2009yt,Blas:2009yd,Blas:2009qj} for the Ho\v{r}ava-Lifshitz gravity and \cite{Gao:2014ula,Gao:2015xwa, Noumi:2016eos} for the massive and bi-gravity), our construction corresponds to a new class of healthy scalar-tensor theories with higher derivatives.

The organization of this paper is as follows. In the next section, after
briefly giving our idea, we will give concrete expressions of the
Horndeski theory in the spatial gauge as an example illuminating our
idea. A complicated part of this result and another interesting
example with higher order derivative terms are given in Appendix \ref{sec:L5} and \ref{appB}.  In
the section \ref{sec:General}, the general framework of our theory will
be given. In the section \ref{sec:Hamiltonian}, we will make Hamiltonian
analysis for our theory and count the number of the degrees of freedom,
which yields only three (two tensors and one scalar) dynamical degree of
freedom without Ostrogradsky's ghosts. Final section is devoted to the conclusion.

\section{Spatial gauge}
\label{sec:Spatial}

\subsection{Spatial gauge}
\label{subsec:Spatial}

\begin{figure}[h]
	\begin{center}
	\includegraphics[scale=0.5]{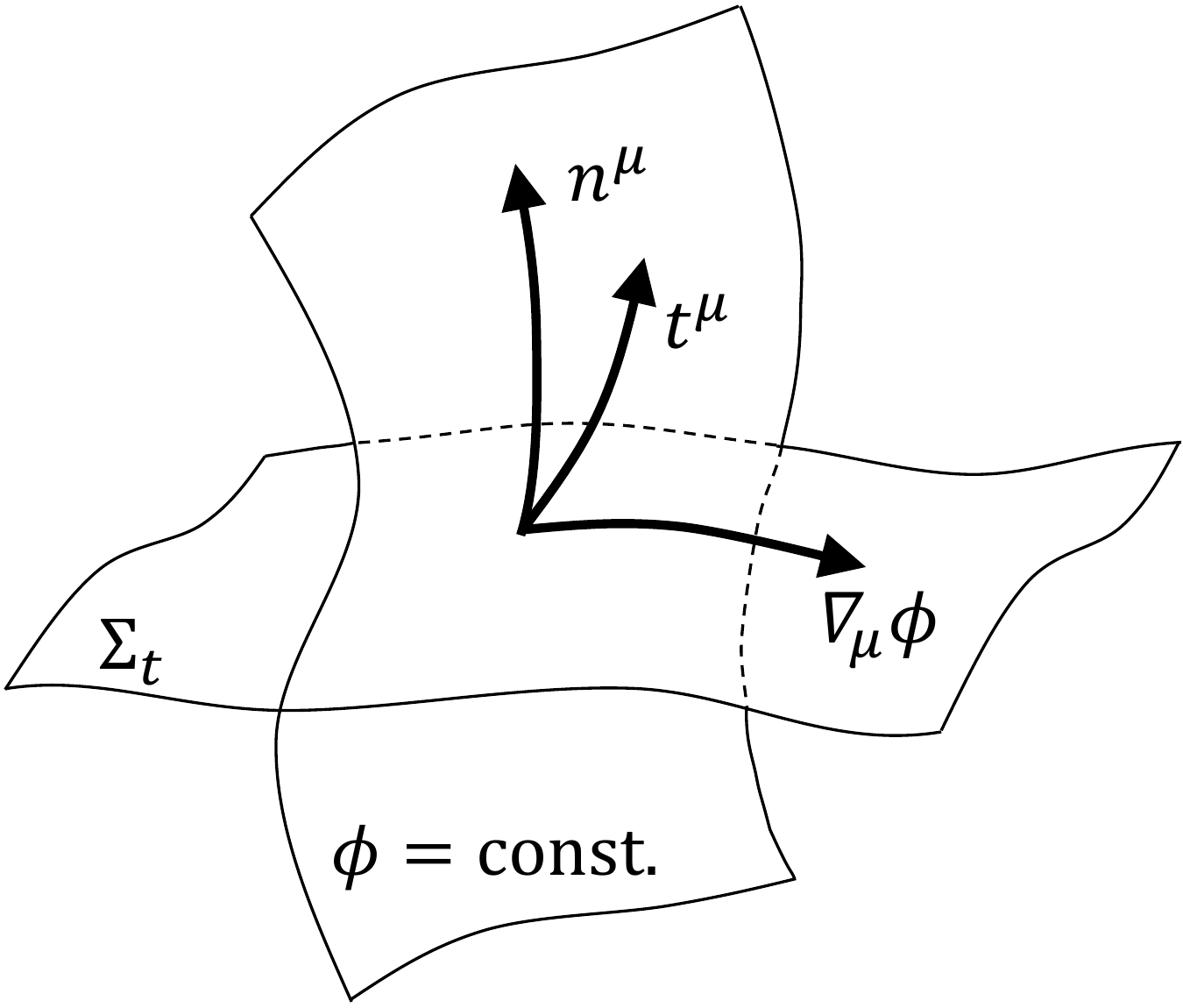}
	\caption{Illustration of the basic picture of the spatial gauge. The vertical and horizonal hypersurfaces represent the constant $\phi$ hypersurfaces and the spatial sector in our 3+1 decomposition, respectively.}
\label{fig:sg}
	\end{center}
\end{figure}

In order to clarify our idea, we assume that the scalar field
$\phi$ possesses a spacelike gradient $\nabla_{\mu}\phi$ that is
non-vanishing everywhere in spacetime.
The basic picture is depicted in Fig.\ref{fig:sg}.
Since $\nabla_{\mu}\phi$ is spacelike, it is possible to choose a set of
timelike curves on the constant $\phi$ hypersurfaces, of which the
tangent vector is $n^{\mu}$.  By definition the Lie derivative of $\phi$
along $n^{\mu}$ is vanishing
\begin{equation}
		\pounds_{\bm{n}} \phi \equiv 0. \label{sg}
\end{equation}
Without loss of generality, we can normalize $n^{\mu}$ so that $n_{\mu}
n^{\mu} = -1$.  We assume there are spacelike hypersurfaces $\Sigma_{t}$
which are orthogonal to $n^{\mu}$.  We now fix the coordinates.
We use $t$ as a time coordinate and $x^{i}$ as arbitrary spatial
coordinates on each foliation $\Sigma_{t}$.
Thus, fixing on the time coordinate through $n^{\mu}$ effectively
breaks time diffeomorphism invariance, which leaves only spatial
diffeomorphism invariance in our theory.
We refer the induced metric on $\Sigma_t$ to $h_{\mu\nu}$ which is
defined through
\begin{equation}
	h_{\mu \nu} := g_{\mu\nu} + n_{\mu} n_{\nu}.
\end{equation}
As usual, we define the lapse and shift functions through the
decomposition of the time vector $t^{\mu} := (\partial_{t})^{\mu}$,
\begin{equation}
	t^{\mu} := N n^{\mu} + N^{\mu}.
\end{equation}
Due to the ambiguity of the choice of spatial coordinates, the time
vector $t^{\mu}$ has not yet been specified at this moment, which may or
may not be parallel to the constant $\phi$ hypersurfaces depending on
the shift functions.

We refer to the above choice of time slices $\Sigma_{t}$ and in particular the condition (\ref{sg}) as the \emph{spatial gauge}. Note the gradient of the scalar field $\nabla_{\mu}\phi$ is split into two parts: the timelike part $\pounds_{\bm{n}} \phi$ and the spacelike part $\mathrm{D}_{\mu}\phi := h_{\mu}^{\phantom{\mu}\nu} \nabla_{\nu}\phi$. If the scalar field is timelike, one is able to fix $\mathrm{D}_{\mu}\phi \equiv 0$, which is dubbed as the unitary gauge in the literature. Thus the spatial gauge is the natural counterpart to the unitary gauge when the scalar field is spacelike.

\subsection{Horndeski theory in the spatial gauge}
\label{subsec:Horndeski}

In order to find the general form of healthy theories in the spatial
gauge, it is useful to derive the gauge-fixed form of a healthy
covariant theory in the spatial gauge.  The starting point is the
Horndeski theory \cite{Horndeski:1974wa}, which describes the most
general single-field scalar-tensor theory with second-order equations of
motion in four-dimensional spacetime. The action is reformulated in
\cite{Deffayet:2011gz,Kobayashi:2011nu}
	\begin{equation}
	\int\mathrm{d}^{4}x\sqrt{-g}\mathcal{L}^{\text{H}}=\int\mathrm{d}^{4}x\sqrt{-g}\left(\mathcal{L}_{2}^{\text{H}}+\mathcal{L}_{3}^{\text{H}}+\mathcal{L}_{4}^{\text{H}}+\mathcal{L}_{5}^{\text{H}}\right),\label{Horndeski}
	\end{equation}
	with
	\begin{eqnarray}
	\mathcal{L}_{2}^{\text{H}} & = & G_{2}\left(X,\phi\right),\label{Horndeski_L2}\\
	\mathcal{L}_{3}^{\text{H}} & = & G_{3}\left(X,\phi\right)\square\phi,\label{Horndeski_L3}\\
	\mathcal{L}_{4}^{\text{H}} & = & G_{4}\left(X,\phi\right)\,{}^{4}\!R+\frac{\partial G_{4}}{\partial X}\left[\left(\square\phi\right)^{2}-\left(\nabla_{\mu}\nabla_{\nu}\phi\right)^{2}\right],\label{Horndeski_L4}\\
	\mathcal{L}_{5}^{\text{H}} & = & G_{5}\left(X,\phi\right)\,{}^{4}\!G^{\mu\nu}\nabla_{\mu}\nabla_{\nu}\phi-\frac{1}{6}\frac{\partial G_{5}}{\partial X}\left[\left(\square\phi\right)^{3}-3\square\phi\left(\nabla_{\mu}\nabla_{\nu}\phi\right)^{2}+2\left(\nabla_{\mu}\nabla_{\nu}\phi\right)^{3}\right],\label{Horndeski_L5}
	\end{eqnarray}
where 
	\begin{equation}
	X\equiv-\frac{1}{2}\left(\nabla\phi\right)^{2},\label{X_def}
	\end{equation}
and $\square\phi \equiv \nabla_{\mu}\nabla^{\mu}\phi$, ${}^{4}\!R$ and
${}^{4}\!G_{\mu\nu}$ are the four-dimensional Ricci scalar and Einstein
tensor, respectively.

Our procedure composes of two steps. First we perform the standard 3+1
decomposition with spatial hypersurfaces specified by $n^{\mu}$. The
basic relations are
	\begin{equation}
	\nabla_{\mu}\phi=-n_{\mu}\pounds_{\bm{n}}\phi+\mathrm{D}_{\mu}\phi,\label{nabla_1_phi_dec}
	\end{equation}
and
	\begin{eqnarray}
	\nabla_{\mu}\nabla_{\nu}\phi & = & n_{\mu}n_{\nu}\left(\pounds_{\bm{n}}^{2}\phi-a^{\rho}\,\mathrm{D}_{\rho}\phi\right)\nonumber \\
	&  & -2n_{(\mu}\left(\mathrm{D}_{\nu)}\pounds_{\bm{n}}\phi-K_{\nu)}^{\phantom{\nu}\rho}\,\mathrm{D}_{\rho}\phi\right)-\pounds_{\bm{n}}\phi\,K_{\mu\nu}+\mathrm{D}_{\mu}\mathrm{D}_{\nu}\phi,\label{nabla_2_phi_dec}
	\end{eqnarray}
as well as the Gauss-Codazzi relations.  Here $\mathrm{D}_{\mu}$ stands
for the covariant derivative with respect to the induced metric
$h_{\mu\nu}$.  In (\ref{nabla_2_phi_dec}), $\mathrm{D}_{\mu}\phi
\equiv h_{\mu}{}^{\nu}\nabla_{\nu}\phi$, $\pounds_{\bm{n}}\phi \equiv
n^{\mu}\nabla_{\mu}\phi$, the acceleration $a_{\mu}$, and the extrinsic
curvature $K_{\mu\nu}$ are introduced by
\begin{align}
 \nabla_{\mu} n_{\nu} = - n_{\mu} a_{\nu} + K_{\mu\nu},
\end{align}
where $a_{\mu} = n^{\nu}\nabla_{\nu}n_{\mu}$. 
Then we take the spatial gauge (i.e., coordinates) by choosing
$\pounds_{\bm{n}}\phi=0$.  It immediately follows that in the spatial
gauge, terms involving $\pounds_{\bm{n}}\phi$ and its spatial
derivatives drop out and we are left with
	\begin{equation}
	 \nabla_{\mu}\phi\rightarrow\delta_{\mu}^{i}\mathrm{D}_{i}\phi,\label{nablaphi_sg}
	\end{equation}
and 
	\begin{equation}
	\nabla_{\mu}\nabla_{\nu}\phi\rightarrow-N^{2}\delta_{\mu}^{0}\delta_{\nu}^{0}a^{i}\mathrm{D}_{i}\phi-2N\delta_{(\mu}^{0}\delta_{\nu)}^{i}K_{i}^{\phantom{i}j}\,\mathrm{D}_{j}\phi+\delta_{\mu}^{i}\delta_{\nu}^{j}\mathrm{D}_{i}\mathrm{D}_{j}\phi.\label{nabla2phi_sg}
	\end{equation}
As the result, the canonical kinetic term $X$ defined in (\ref{X_def})
reduces to
	\begin{equation}
	X\rightarrow-\frac{1}{2}\mathrm{D}_{i}\phi\mathrm{D}^{i}\phi. \label{X_sg}
	\end{equation}
In the above and the following, $\mathrm{D}_{i}$ is the covariant derivative adapted to $h_{ij}$. The indices $i,j, \dots$ stand for the components with respect to the basis
\begin{equation}
 \omega^{i}  := N^{i} \mathrm{d}t + \mathrm{d}x^i,
\end{equation} 
and these indices are raised and lowered by $h^{ij}$ and $h_{ij}$.
For example, $\mathrm{D}_i \phi$ is defined by $\mathrm{D}_{\mu} \phi \, \mathrm{d}x^{\mu} = \mathrm{D}_{i} \phi \, \omega^i$.

In the spatial gauge, $\mathcal{L}_{2}^{\mathrm{H,(s.g.)}}$  takes the same functional form as (\ref{Horndeski_L2})  but with $X$ replaced by (\ref{X_sg}).
After some manipulations, we find
	\begin{equation}
	\mathcal{L}_{3}^{\mathrm{H,(s.g.)}}=\frac{\partial G_{3}}{\partial X}\mathrm{D}_{i}\mathrm{D}_{j}\phi\mathrm{D}^{i}\phi\mathrm{D}^{j}\phi-\frac{\partial G_{3}}{\partial\phi}\mathrm{D}_{i}\phi\mathrm{D}^{i}\phi.\label{L2H_sg}
	\end{equation}
and 
	\begin{eqnarray}
	\mathcal{L}_{4}^{\mathrm{H,(s.g.)}} & = & G_{4}\left(R+K_{ij}K^{ij}-K^{2}\right)+2\frac{\partial G_{4}}{\partial X}\left(R^{ij}+K^{ik}K_{k}^{\phantom{k}j}-KK^{ij}\right)\mathrm{D}_{i}\phi\mathrm{D}_{j}\phi\nonumber \\
	&  & -\frac{\partial G_{4}}{\partial X}\left[\left(\mathrm{D}^{2}\phi\right)^{2}-\mathrm{D}_{i}\mathrm{D}_{j}\phi\mathrm{D}^{i}\mathrm{D}^{j}\phi\right]\nonumber \\
	&  & +2\frac{\partial^{2}G_{4}}{\partial X^{2}}\left(\mathrm{D}_{i}\mathrm{D}_{j}\phi\mathrm{D}^{i}\phi\mathrm{D}^{j}\phi\mathrm{D}^{2}\phi-\mathrm{D}^{i}\mathrm{D}_{j}\phi\mathrm{D}^{j}\phi\mathrm{D}_{i}\mathrm{D}_{k}\phi\mathrm{D}^{k}\phi\right)\nonumber \\
	&  & +2\frac{\partial^{2}G_{4}}{\partial X\partial\phi}\left(2\mathrm{D}_{i}\mathrm{D}_{j}\phi\mathrm{D}^{i}\phi\mathrm{D}^{j}\phi-\mathrm{D}^{i}\phi\mathrm{D}_{i}\phi\mathrm{D}^{2}\phi\right)\nonumber \\
	&  & -2\frac{\partial^{2}G_{4}}{\partial\phi^{2}}\mathrm{D}_{i}\phi\mathrm{D}^{i}\phi-2\frac{\partial G_{4}}{\partial\phi}\mathrm{D}^{2}\phi.\label{L4H_sg}
	\end{eqnarray}
The cubic Horndeski Lagrangian in the spatial gauge
$\mathcal{L}_{5}^{\mathrm{H},\mathrm{(s.g.)}}$ can be found in
(\ref{L5H_sg}) due to its length.

When the scalar field possesses a timelike gradient, the 3+1
decomposition of the Horndeski theory in the unitary gauge has been
given in \cite{Gleyzes:2013ooa} (see also \cite{Fujita:2015ymn}). It is
clear that the above expressions are complementary  to the expressions in
the unitary gauge.

\section{General framework}
\label{sec:General}

The Horndeski theory is a generally covariant scalar-tensor theory,
which is healthy no matter the scalar field is timelike or spacelike.
In other words, the Lagrangians in the spatial gauge
$\mathcal{L}_{2}^{\mathrm{H,(s.g.)}}$,
$\mathcal{L}_{3}^{\mathrm{H,(s.g.)}}$, $\mathcal{L}_{4}^{\mathrm{H,(s.g.)}}$
and $\mathcal{L}_{5}^{\mathrm{H,(s.g.)}}$ represent explicit examples for
healthy scalar-tensor theories in the spatial gauge.  We have several
observations from the expressions of
$\mathcal{L}_{2}^{\mathrm{H,(s.g.)}}$,
$\mathcal{L}_{3}^{\mathrm{H,(s.g.)}}$, $\mathcal{L}_{4}^{\mathrm{H,(s.g.)}}$
and $\mathcal{L}_{5}^{\mathrm{H,(s.g.)}}$.  First, in the spatial gauge,
instead of an overall factor, the lapse function $N$ enters the
Lagrangians only in terms of the extrinsic curvature.  Second, and most importantly, the
action explicitly depends on a non-dynamical scalar field $\phi$, which
has no time derivative terms, while spatial derivative terms such as
$\mathrm{D}_{i}\phi$, $\mathrm{D}_{i}\mathrm{D}_{j}\phi$ etc are
generally allowed.

Inspired by the above observations, we consider the following general action
\begin{equation}
S^{\mathrm{(s.g.)}}=\int\mathrm{d}t\mathrm{d}^{3}x\,N\sqrt{h}\,\mathcal{L}\left(h_{ij},K_{ij},R_{ij},N,\phi,\mathrm{D}_{i} \right).\label{S_ori}
\end{equation}
By inclusion of the nonlinear dependence of $N$, \eqref{S_ori} includes a very general class of higher derivative scalar-tensor theories, which is beyond the DHOST theory \cite{Langlois:2015cwa,Langlois:2015skt,Achour:2016rkg} in the spatial gauge (see Appendix \ref{appB}). One may worry that our theory possibly includes 4 degrees of freedom, one of which is the Ostrogradsky's ghost due to the higher derivatives.
One of the main purposes of this work is to show that the model (\ref{S_ori}) propagates up to 3 degrees of freedom, through a detailed Hamiltonian constraint analysis.

We have three comments on the action (\ref{S_ori}). First, once we write down
this action, one can extend the theory by regarding that the scalar field can be not only spacelike but also
timelike. We have only to write down an action on some spacelike
hypersurface. Second, since the action does not have a kinetic term (no
time derivative) for the scalar field, the scalar field itself is non-dynamical.
Nevertheless, an alternative dynamical scalar degree of freedom arises due to the explicit breaking of general covariance in the action (\ref{S_ori}).  Finally,
the action (\ref{S_ori}) is invariant under spatial diffeomorphism as
explained in the previous section. The timelike vector $n^{\mu}$
orthogonal to some spacelike hypersurface fixes time slice and breaks
time diffeomorphism invariance while keeping spatial one. In this
sense, it can be viewed as a novel class of spatially covariant theories of
gravity, and can be compared with action in
\cite{Gao:2014soa,Gao:2014fra} (see also \cite{Gao:2018znj}) generally
given by in the comoving gauge $\phi = t$ (a special case of the
unitary gauge),
\begin{equation}
	S^{\mathrm{(c.g.)}}=\int\mathrm{d}t\mathrm{d}^{3}x\,N\sqrt{h}\,\mathcal{L}\left(t,h_{ij},K_{ij},R_{ij},N,\mathrm{D}_{i} \right).\label{S_com}
\end{equation}
We would like to emphasize that our theory \eqref{S_ori} does not
reduce to \eqref{S_com} even if one allows timelike $\phi$ and imposes
an ansatz $\phi(t) = t$, since the constraint equation from the $\phi$
variation is not imposed in \eqref{S_com}. To be precise, even if
the scalar field $\phi$ is timelike, since the action in (\ref{S_ori}) has no
general covariance, it does not admit gauge freedom to choose $\phi =
\phi(t)$.  One may also worry that our theory \eqref{S_ori} reduces to
\eqref{S_com}, except for explicit $t$ dependence, after integrating
$\phi$ out. However this is not true because our theory \eqref{S_ori}
generally include spatial derivative of $\phi$ and inverse of spatial
derivatives must be included after eliminating $\phi$. 

For our purpose, it is convenient to rewrite (\ref{S_ori}) in an equivalent form
	\begin{equation}
	\tilde{S} := \int \mathrm{d}t \mathrm{d}^3 x N \sqrt{h} \tilde{{\cal L}} :=S+\int\mathrm{d}t\mathrm{d}^{3}x\frac{\delta S}{\delta B_{ij}}\left(K_{ij}-B_{ij}\right),\label{S_tld_xpl}
	\end{equation}
where $S$ is the same action as (\ref{S_ori}) with $K_{ij}$ replaced by
$B_{ij}$, i.e.,
	\begin{equation}
	S:=\int\mathrm{d}t\mathrm{d}^{3}x\,N\sqrt{h}\,\mathcal{L}\left(h_{ij},B_{ij},R_{ij},N,\phi,\mathrm{D}_{i}\right).\label{S_def}
	\end{equation}
Note that the lapse function $N$ appears \emph{not} merely as the overall factor
in the above action.

At this point, note $\tilde{S}$ depends on 17 variables:
	\begin{equation}
	\Phi_{I}:=\left\{ N^{i},\phi,B_{ij},N,h_{ij}\right\} ,\label{var}
	\end{equation}
where the indices $I,J,\cdots$ formally denote different kinds of
variables as well as their tensorial indices.

\section{Hamiltonian analysis}
\label{sec:Hamiltonian}

\subsection{Primary constraints}
\label{subsec:Primary}

The conjugate momenta corresponding to the variables (\ref{var}) are defined as
	\begin{equation}
		\Pi^{I} := \frac{\delta \tilde{S}}{\delta\dot{\Phi}_{I}}, \label{mom_def}
	\end{equation}
which are explicitly given by
	\begin{eqnarray}
	\pi_{i} & := & \frac{\delta\tilde{S}}{\delta\dot{N}^{i}}=0,\label{mom_1}\\
	p & := & \frac{\delta\tilde{S}}{\delta\dot{\phi}}=0,\qquad p^{ij}:=\frac{\delta\tilde{S}}{\delta\dot{B}_{ij}}=0\label{mom_2}\\
	\pi & := & \frac{\delta\tilde{S}}{\delta\dot{N}}=0,\qquad\pi^{ij}:=\frac{\delta\tilde{S}}{\delta\dot{h}_{ij}}=\frac{1}{2N}\frac{\delta S}{\delta B_{ij}},\label{mom_3}
	\end{eqnarray}
According to (\ref{mom_1})-(\ref{mom_3}), there are in total 17 primary constraints
	\begin{equation}
	\pi_{i}\approx 0,\qquad p\approx 0,\qquad p^{ij}\approx 0,\qquad\pi\approx 0,\qquad\tilde{\pi}^{ij}:=\pi^{ij}-\frac{1}{2N}\frac{\delta S}{\delta B_{ij}}\approx 0,\label{pc}
	\end{equation}
where and throughout this work ``$\approx$'' represents the so-called ``weak equality'' that holds when the primary constraints are taken into account. For later convenience, we denote
	\begin{equation}
	\Pi^{I}=\left\{ \pi_{i},p,p^{ij},\pi,\pi^{ij}\right\} ,\label{mom}
	\end{equation}
for the set of momenta, and
	\begin{equation}
	\varphi^{I}:=\left\{ \pi_{i},p,p^{ij},\pi,\tilde{\pi}^{ij}\right\}, \label{pri_cons}
	\end{equation}
for the set of primary constraints.

\subsection{Canonical Hamiltonian}
\label{subsec:Canonical}

The canonical Hamiltonian is obtained by 
	\begin{equation}
	H_{\mathrm{C}}:=\int\mathrm{d}^{3}x\left(\sum_{I}\Pi^{I}\dot{\Phi}_{I}-N\sqrt{h}\tilde{\mathcal{L}}\right)\approx\int\mathrm{d}^{3}x\left(NC+\pi^{ij}\pounds_{\vec{N}}\,h_{ij}\right),\label{Ham_C_def}
	\end{equation}
with
	\begin{equation}
	C:=2\pi^{ij}B_{ij}-\sqrt{h}\mathcal{L}.\label{C_def}
	\end{equation}
The definition of the canonical Hamiltonian includes ambiguity of how to use the primary constraints. By the use of this
ambiguity, one can simplify the following constraint analysis. One of
the crucial properties of our model (\ref{S_ori}) is that it possesses
spatial covariance. As a result, techniques that are used in the
Hamiltonian analysis for the spatially covariant gravity in
\cite{Gao:2018znj} can be also applied.  Especially, inspired by the
discussion in \cite{Gao:2018znj}, it is convenient to define the
``proper'' canonical Hamiltonian to be
	\begin{equation}
	H_{\mathrm{C}} :=\int\mathrm{d}^{3}x\left(NC\right)+X[\vec{N}],\label{Ham_C}
	\end{equation}
where $C$ is given in (\ref{C_def}), and for a general spatial vector
field $\vec{\xi}$, $X[\vec{\xi}]$ is defined to be
	\begin{equation}
	X[\vec{\xi}]:=\int\mathrm{d}^{3}x\sum_{I}\Pi^{I}\pounds_{\vec{\xi}}\,\Phi_{I}\simeq\int\mathrm{d}^{3}x\,\xi^{i}\mathcal{C}_{i},\label{X_xi_ibp}
	\end{equation}
with
	\begin{eqnarray}
	\mathcal{C}_{i} & = & \pi\mathrm{D}_{i}N-2\sqrt{h}\mathrm{D}_{j}\left(\frac{1}{\sqrt{h}}\pi_{i}^{j}\right)+p\mathrm{D}_{i}\phi+p^{kl}\mathrm{D}_{i}B_{kl}-2\sqrt{h}\mathrm{D}_{j}\left(\frac{p^{jk}}{\sqrt{h}}B_{ik}\right)\nonumber \\
	&  & +\pi_{j}\mathrm{D}_{i}N^{j}+\sqrt{h}\mathrm{D}_{j}\left(\frac{1}{\sqrt{h}}\pi_{i}N^{j}\right).\label{calCi_def}
	\end{eqnarray}
The canonical Hamiltonian $H_{\mathrm{C}}$ defined in (\ref{Ham_C})
weakly equals to (\ref{Ham_C_def}).  We prefer to use (\ref{Ham_C}) due
to the following property of $X[\vec{\xi}]$. For any functional on the
phase space $\mathcal{F} = \mathcal{F}[\Phi_{I},\Pi^{I}]$ that is
invariant under the time-independent spatial diffeomorphism, the
following equality holds
	\begin{equation}
	\left[\int\mathrm{d}^{3}x\sum_{I}\Pi^{I}\pounds_{\vec{\xi}}\,\Phi_{I},\mathcal{F}\right]=\int\mathrm{d}^{3}x\sum_{I}\Pi^{I}\pounds_{[\vec{\xi},\mathcal{F}]}\Phi_{I},\label{PB_LD_xi}
	\end{equation}
or compactly
	\begin{equation}
	\left[X[\vec{\xi}],\mathcal{F}\right]=X\left[[\vec{\xi},\mathcal{F}]\right],\label{PB_LD_xi_com}
	\end{equation}
up to a boundary term. We refer to \cite{Gao:2018znj} for more details
and for the general proof.  As we shall see, defining the canonical
Hamiltonian as (\ref{Ham_C}) significantly simplifies the calculation of
Poisson brackets.

Due to the presence of primary constraints, the time-evolution is
determined by the so-called total Hamiltonian, which is given by
\begin{equation}
H_{\mathrm{T}} := H_{\mathrm{C}}+\int\mathrm{d}^{3}y\sum_{I}\lambda_{I}\left(\vec{y}\right)\varphi^{I}\left(\vec{y}\right),\label{Ham_total}
\end{equation}
where $\lambda_{I}:= 
\{v^{i}, v, v_{ij}, \lambda, \lambda_{ij} \}$ are undetermined Lagrange
multipliers which are associated with $\varphi^{I} = \{\pi_{i}, p,
p^{ij}, \pi, \tilde{\pi}^{ij}\}$.

\subsection{Constraint algebra}
\label{subsec:Constraint}

The matrix composed of Poisson brackets of the primary constraints takes
the following form
	\begin{eqnarray}
	&  & \left[\varphi^{I}\left(\vec{x}\right),\varphi^{J}\left(\vec{y}\right)\right]\nonumber \\
	& = & \left(\begin{array}{ccccc}
	0 & 0 & 0 & 0 & 0\\
	0 & 0 & 0 & 0 & \left[p\left(\vec{x}\right),\tilde{\pi}^{kl}\left(\vec{y}\right)\right]\\
	0 & 0 & 0 & 0 & \left[p^{ij}\left(\vec{x}\right),\tilde{\pi}^{kl}\left(\vec{y}\right)\right]\\
	0 & 0 & 0 & 0 & \left[\pi\left(\vec{x}\right),\tilde{\pi}^{kl}\left(\vec{y}\right)\right]\\
	0 & \left[\tilde{\pi}^{ij}\left(\vec{x}\right),p\left(\vec{y}\right)\right] & \left[\tilde{\pi}^{ij}\left(\vec{x}\right),p^{kl}\left(\vec{y}\right)\right] & \left[\tilde{\pi}^{ij}\left(\vec{x}\right),\pi\left(\vec{y}\right)\right] & \left[\tilde{\pi}^{ij}\left(\vec{x}\right),\tilde{\pi}^{kl}\left(\vec{y}\right)\right]
	\end{array}\right),\label{PB_pc}
	\end{eqnarray}
where the generally non-vanishing entries are 
	\begin{eqnarray}
	\left[p\left(\vec{x}\right),\tilde{\pi}^{kl}\left(\vec{y}\right)\right] & = & \frac{1}{2N\left(\vec{y}\right)}\frac{\delta^{2}S}{\delta\phi\left(\vec{x}\right)\delta B_{kl}\left(\vec{y}\right)},\label{PB_p_pitldkl}\\
	\left[p^{ij}\left(\vec{x}\right),\tilde{\pi}^{kl}\left(\vec{y}\right)\right] & = & \frac{1}{2}\frac{1}{N\left(\vec{y}\right)}\frac{\delta^{2}S}{\delta B_{ij}\left(\vec{x}\right)\delta B_{kl}\left(\vec{y}\right)},\label{PB_pij_pitldkl}\\
	\left[\pi\left(\vec{x}\right),\tilde{\pi}^{ij}\left(\vec{y}\right)\right] & = & -\frac{1}{2}\delta^{3}\left(\vec{x}-\vec{y}\right)\frac{1}{N^{2}\left(\vec{y}\right)}\frac{\delta S}{\delta B_{ij}\left(\vec{y}\right)}+\frac{1}{2}\frac{1}{N\left(\vec{y}\right)}\frac{\delta^{2}S}{\delta N\left(\vec{x}\right)\delta B_{ij}\left(\vec{y}\right)},\label{PB_pi_pitldij}\\
	\left[\tilde{\pi}^{ij}\left(\vec{x}\right),\tilde{\pi}^{kl}\left(\vec{y}\right)\right] & = & \frac{1}{2N\left(\vec{y}\right)}\frac{\delta^{2}S}{\delta h_{ij}\left(\vec{x}\right)\delta B_{kl}\left(\vec{y}\right)}-\frac{1}{2N\left(\vec{x}\right)}\frac{\delta^{2}S}{\delta B_{ij}\left(\vec{x}\right)\delta h_{kl}\left(\vec{y}\right)}.\label{PB_pitldij_pitldkl}
	\end{eqnarray}
Before we proceed, note if the lapse function $N$ enters
(\ref{S_def}) linearly as in the case of Horndeski theory, we may evaluate (\ref{PB_pi_pitldij}) explicitly and find
	\begin{eqnarray}
	\left[\pi\left(\vec{x}\right),\tilde{\pi}^{ij}\left(\vec{y}\right)\right] & = & -\frac{1}{2}\delta^{3}\left(\vec{x}-\vec{y}\right)\frac{1}{N}\sqrt{h}\sum_{n=1}\left(-1\right)^{n}\partial_{k_{1}}\cdots\partial_{k_{n}}\frac{\partial\mathcal{L}}{\partial\left(\partial_{k_{1}}\cdots\partial_{k_{n}}B_{ij}\right)}\nonumber \\
	&  & +\frac{1}{2}\frac{\sqrt{h\left(\vec{x}\right)}}{N\left(\vec{y}\right)}\sum_{n=1}\left(-1\right)^{n}\partial_{y^{k_{1}}}\cdots\partial_{y^{k_{n}}}\delta^{3}\left(\vec{x}-\vec{y}\right)\frac{\partial\mathcal{L}\left(\vec{x}\right)}{\partial\left(\partial_{x^{k_{1}}}\cdots\partial_{x^{k_{n}}}B_{ij}\right)}.
	\end{eqnarray}
Then
$\left[\pi\left(\vec{x}\right),\tilde{\pi}^{ij}\left(\vec{y}\right)\right]\neq
0$ only if our Lagrangian in (\ref{S_def}) depends on spatial
derivatives of $B_{ij}$, i.e., the extrinsic curvature $K_{ij}$ enters
the original action (\ref{S_ori}) with spatial derivatives.
Fortunately, the following counting number of degrees of freedom is not
subject to this subtlety.

Constraints must be preserved in time. For the primary constraints $\varphi^{I}$ in (\ref{pri_cons}), we must require that
	\begin{equation}
	\frac{\mathrm{d}\varphi^{I}\left(\vec{x}\right)}{\mathrm{d}t}=\int\mathrm{d}^{3}y\sum_{J}\left[\varphi^{I}\left(\vec{x}\right),\varphi^{J}\left(\vec{y}\right)\right]\lambda_{J}\left(\vec{y}\right)+\left[\varphi^{I}\left(\vec{x}\right),H_{\mathrm{C}}\right], \label{cc_pc}
	\end{equation}
where $\left[\varphi^{I}\left(\vec{x}\right),\varphi^{J}\left(\vec{y}\right)\right]$ is given in (\ref{PB_pc}). 
After some manipulations, we find
	\begin{equation}
	\left[\varphi^{I}\left(\vec{x}\right),H_{\mathrm{C}}\right]=\left(\begin{array}{c}
	-\mathcal{C}_{i}\left(\vec{x}\right)\\
	\frac{\delta S}{\delta\phi\left(\vec{x}\right)}\\
	0\\
	-\mathcal{C}\left(\vec{x}\right)\\
	\left[\tilde{\pi}^{ij}\left(\vec{x}\right),H_{\mathrm{C}}\right]
	\end{array}\right),
	\end{equation}
where $\mathcal{C}_{i}$ is defined in (\ref{calCi_def}), $\mathcal{C}$ is given by
\begin{align}
 \mathcal{C}(\vec{x}) &= C(\vec{x}) + \int \mathrm{d}^3 y N(\vec{y}) \frac{\delta C(\vec{y})}{\delta N(\vec{x})}, \notag\\
 &= 2 \pi^{ij}(\vec{x}) B_{ij}(\vec{x}) - \frac{\delta S}{\delta N(\vec{x})},
\end{align}
 and
	\begin{equation}
	\left[\tilde{\pi}^{ij}\left(\vec{x}\right),H_{\mathrm{C}}\right]=\frac{\delta S}{\delta h_{ij}\left(\vec{x}\right)}-\frac{1}{N\left(\vec{x}\right)}\int\mathrm{d}^{3}z\frac{\delta^{2}S}{\delta B_{ij}\left(\vec{x}\right)\delta h_{kl}\left(\vec{z}\right)}N\left(\vec{z}\right)B_{kl}\left(\vec{z}\right).\label{PB_pi^ij_tld_HC}
	\end{equation}

The consistency conditions (\ref{cc_pc}) yield the following 5
equations:
	\begin{eqnarray}
	-\mathcal{C}_{i}\left(\vec{x}\right)  &=&  0,\label{cc_pri_1}\\
	\int\mathrm{d}^{3}y\left[p\left(\vec{x}\right),\tilde{\pi}^{kl}\left(\vec{y}\right)\right]\lambda_{kl}\left(\vec{y}\right)+\frac{\delta S}{\delta\phi\left(\vec{x}\right)} &=&  0,\label{cc_pri_2}\\
	\int\mathrm{d}^{3}y\left[p^{ij}\left(\vec{x}\right),\tilde{\pi}^{kl}\left(\vec{y}\right)\right]\lambda_{kl}\left(\vec{y}\right)  &=&  0,\label{cc_pri_3}\\
	\int\mathrm{d}^{3}y\left[\pi\left(\vec{x}\right),\tilde{\pi}^{kl}\left(\vec{y}\right)\right]\lambda_{kl}\left(\vec{y}\right) - \mathcal{C}\left(\vec{x}\right)  &=&  0,\label{cc_pri_4}\\
	\int\mathrm{d}^{3}y\Big\{\left[\tilde{\pi}^{ij}\left(\vec{x}\right),p\left(\vec{y}\right)\right]v\left(\vec{y}\right)+\left[\tilde{\pi}^{ij}\left(\vec{x}\right),p^{kl}\left(\vec{y}\right)\right]v_{kl}\left(\vec{y}\right)&&\nonumber \\
	\qquad +\left[\tilde{\pi}^{ij}\left(\vec{x}\right),\pi\left(\vec{y}\right)\right]\lambda\left(\vec{y}\right)+\left[\tilde{\pi}^{ij}\left(\vec{x}\right),\tilde{\pi}^{kl}\left(\vec{y}\right)\right]\lambda_{kl}\left(\vec{y}\right)\Big\}+\left[\tilde{\pi}^{ij}\left(\vec{x}\right),H_{\mathrm{C}}\right] & =&  0, \label{cc_pri_5}
	\end{eqnarray}
where Poisson brackets among the primary constraints are given in
(\ref{PB_p_pitldkl})-(\ref{PB_pitldij_pitldkl}).  It immediately follows
from (\ref{cc_pri_1}) that
	\begin{equation}
	\mathcal{C}_{i}\approx 0,
	\end{equation}
are 3 secondary constraints, as expected. Since we assume
$\frac{\delta^{2}S}{\delta B_{ij}\left(\vec{x}\right)\delta
B_{kl}\left(\vec{y}\right)}$ is not degenerate, we are able to fix the
Lagrange multiplies from (\ref{cc_pri_3}) to be
\begin{equation}
\lambda_{ij}=0.\label{lambda_ij_sol}
\end{equation}
Then (\ref{cc_pri_2}) and (\ref{cc_pri_4}) yield another two secondary constraints:
\begin{equation}
\frac{\delta S}{\delta\phi}\approx 0,\qquad \mathcal{C}\approx 0.
\end{equation}
Note since $\phi$ itself has no dynamics, its equation of motion
$\frac{\delta S}{\delta\phi}=0$ must be a constraint, as expected.  The
last equation (\ref{cc_pri_5}) simply further fixes the Lagrange
multiplier $v_{ij}$ instead of generating new constraint. To summarize,
the time consistency equations for the 17 primary constraints reduces to
5 secondary constraints $\mathcal{C}_i, \mathcal{C}$ and $\delta S/ \delta \phi$ and 12
equations which fix the Lagrange multipliers $\lambda_{ij}$ and
$v_{ij}$.

We also need to check the consistency conditions for the secondary
constraints got so far.  To this end, first we evaluate the Poisson
brackets of the secondary constraints with the primary constraints as
well as with themselves.  Thanks to the property (\ref{PB_LD_xi_com}),
one can show that $\mathcal{C}_{i}$ has vanishing Poisson bracket with all the constraints (see \cite{Gao:2018znj} as well as Appendix \ref{sec:PBCi} for details).
The Poisson brackets of $\mathcal{C}\approx 0$ with the primary
constraints are:
	\begin{equation}
	\left[\pi_{i}\left(\vec{x}\right),\mathcal{C}\left(\vec{y}\right)\right]=0,
	\end{equation}
and
	\begin{eqnarray}
\left[\pi(\vec{x}), \mathcal{C}(\vec{y})\right] &=& 
\frac{\delta^2 S}{\delta N(\vec{x}) \delta N(\vec{y})} , \\
\left[p\left(\vec{x}\right),\mathcal{C} \left(\vec{y}\right)\right] & = & \frac{\delta^2 S}{\delta \phi(\vec{x}) \delta N(\vec{y})} ,\label{PB_p_C}\\
	\left[p^{ij}\left(\vec{x}\right),\mathcal{C}\left(\vec{y}\right)\right] & = & -2\delta^{3}\left(\vec{x}-\vec{y}\right)\pi^{ij}\left(\vec{y}\right)
+
\frac{\delta^2 S}{\delta B_{ij}(\vec{x}) \delta N(\vec{y})},
\label{PB_p^ij_C}\\
	\left[\tilde{\pi}^{ij}\left(\vec{x}\right), \mathcal{C}\left(\vec{y}\right)\right] & = & \frac{\delta^2 S}{\delta h_{ij}\left(\vec{x}\right) \delta N (\vec{y})}-\frac{1}{N\left(\vec{x}\right)}\frac{\delta^{2}S}{\delta B_{ij}\left(\vec{x}\right)\delta h_{kl}\left(\vec{y}\right)}B_{kl}\left(\vec{y}\right).\label{PB_pitld^ij_C}
	\end{eqnarray}
The Poisson brackets of $\frac{\delta S}{\delta\phi} \approx 0$ with the
primary constraints are:
	\begin{eqnarray}
	\left[\pi_{i}\left(\vec{x}\right),\frac{\delta S}{\delta\phi\left(\vec{y}\right)}\right] & = & 0,\label{PB_pi_i_dSdf}
	\end{eqnarray}
and
	\begin{eqnarray}
	\left[p\left(\vec{x}\right),\frac{\delta S}{\delta\phi\left(\vec{y}\right)}\right] & = & -\frac{\delta^{2}S}{\delta\phi\left(\vec{x}\right)\delta\phi\left(\vec{y}\right)},\label{PB_p_dSdf}\\
	\left[p^{ij}\left(\vec{x}\right),\frac{\delta S}{\delta\phi\left(\vec{y}\right)}\right] & = & -\frac{\delta^{2}S}{\delta B_{ij}\left(\vec{x}\right)\delta\phi\left(\vec{y}\right)},\label{PB_p^ij_dSdf}\\
	\left[\pi\left(\vec{x}\right),\frac{\delta S}{\delta\phi\left(\vec{y}\right)}\right] & = & -\frac{\delta^{2}S}{\delta N\left(\vec{x}\right)\delta\phi\left(\vec{y}\right)},\label{PB_pi_dSdf}\\
	\left[\tilde{\pi}^{ij}\left(\vec{x}\right),\frac{\delta S}{\delta\phi\left(\vec{y}\right)}\right] & = & -\frac{\delta^{2}S}{\delta h_{ij}\left(\vec{x}\right)\delta\phi\left(\vec{y}\right)}.\label{PB_pitld^ij_dSdf}
	\end{eqnarray}
Finally, the Poisson brackets between $\mathcal{C}$ and $\frac{\delta S}{\delta \phi}$ are
	\begin{eqnarray}
	\left[\mathcal{C}(\vec{x}),\mathcal{C}(\vec{y})\right] & = & -  \frac{\delta^2 S}{\delta h_{ij}\left(\vec{y}\right) \delta N(\vec{x})}  2 B_{ij} (\vec{y})  +2B_{ij}\left(\vec{x}\right)  \frac{\delta^2 S}{\delta h_{ij}\left(\vec{x}\right) \delta N(\vec{y})}
,\label{PB_C_C}\\
	\left[\frac{\delta S}{\delta\phi\left(\vec{x}\right)},\frac{\delta S}{\delta\phi\left(\vec{y}\right)}\right] & = & 0,\label{PB_dSdf_dSdf}\\
	\left[\frac{\delta S}{\delta\phi(\vec{x})}, \mathcal{C}(\vec{y})\right] & = & \frac{\delta^{2}S}{\delta\phi(\vec{x})\delta h_{ij}\left(\vec{y}\right)}2B_{ij}\left(\vec{y}\right).\label{PB_dSdf_C}
	\end{eqnarray}

We are now ready to check the consistency conditions for the secondary
constraints.  Since $\mathcal{C}_{i}$ has vanishing Poisson brackets
with all the constraints got so far as well as $H_{\mathrm{C}}$, its
consistency condition is automatically satisfied.  On the other hand,
the consistency condition for $\mathcal{C}\approx 0$ and $\frac{\delta
S}{\delta \phi} \approx 0$ yield two equations:
	\begin{eqnarray}
	0\equiv\frac{\mathrm{d} \mathcal{C} \left(\vec{x}\right)}{\mathrm{d}t} & = & \int\mathrm{d}^{3}y \left[\mathcal{C} \left(\vec{x}\right), \pi \left(\vec{y}\right)\right] \lambda \left(\vec{y}\right) + \int\mathrm{d}^{3}y \left[\mathcal{C} \left(\vec{x}\right),p\left(\vec{y}\right)\right]v\left(\vec{y}\right) \nonumber \\
	&  & + \int\mathrm{d}^{3}y \left[\mathcal{C}\left(\vec{x}\right),p^{ij}\left(\vec{y}\right)\right]v_{ij}\left(\vec{y}\right)+\left[\mathcal{C}\left(\vec{x}\right),H_{\mathrm{C}}\right],\label{cc_C}
	\end{eqnarray}
	and
	\begin{eqnarray}
	0\equiv\frac{\mathrm{d}}{\mathrm{d}t}\left(\frac{\delta S}{\delta\phi\left(\vec{x}\right)}\right) & = &  \int\mathrm{d}^{3}y \left[\frac{\delta S}{\delta\phi\left(\vec{x}\right)},\pi\left(\vec{y}\right)\right]\lambda\left(\vec{y}\right) + \int\mathrm{d}^{3}y \left[\frac{\delta S}{\delta\phi\left(\vec{x}\right)},p\left(\vec{y}\right)\right]v\left(\vec{y}\right)
\nonumber \\
	&  &
+\int\mathrm{d}^{3}y \left[\frac{\delta S}{\delta\phi\left(\vec{x}\right)},p^{ij}\left(\vec{y}\right)\right]v_{ij}\left(\vec{y}\right)
+\left[\frac{\delta S}{\delta\phi\left(\vec{x}\right)},H_{\mathrm{C}}\right],\label{cc_dSdf}
	\end{eqnarray}
where various Poisson brackets are given in
(\ref{PB_p_C})-(\ref{PB_pitld^ij_dSdf}), and
	\begin{eqnarray}
	\left[\mathcal{C}\left(\vec{x}\right),H_{\mathrm{C}}\right] & = & 2B_{ij}\left(\vec{x}\right)\frac{\delta S}{\delta h_{ij}\left(\vec{x}\right)} -\int\mathrm{d}^{3}z \frac{\delta S}{\delta N(\vec{x}) \delta h_{ij}\left(\vec{z}\right) }N\left(\vec{z}\right)2B_{ij}\left(\vec{z}\right),
	\end{eqnarray}
	and
	\begin{eqnarray}
	\left[\frac{\delta S}{\delta\phi\left(\vec{x}\right)},H_{\mathrm{C}}\right] & = & \int\mathrm{d}^{3}z\frac{\delta^{2}S}{\delta\phi\left(\vec{x}\right)\delta h_{ij}\left(\vec{z}\right)}N\left(\vec{z}\right)2B_{ij}\left(\vec{z}\right).
	\end{eqnarray}
The consistency conditions (\ref{cc_C}) and (\ref{cc_dSdf}) (together
with (\ref{cc_pri_5})) merely fix the Lagrange multipliers $v$ and
$\lambda$, provided that the coefficient matrix of $v$ and $\lambda$ is non-singular, i.e.,
\begin{align}
\det
 \begin{pmatrix}
  \frac{\delta^2 S}{ \delta N(\vec{x}) \delta N(\vec{y})} & \frac{\delta^2 S}{ \delta N(\vec{x}) \delta \phi(\vec{y})} \\
  \frac{\delta^2 S}{ \delta \phi(\vec{x}) \delta N(\vec{y})} & \frac{\delta^2 S}{ \delta \phi(\vec{x}) \delta \phi(\vec{y})} 
 \end{pmatrix}
\neq 0,\label{det}
\end{align}
on the constraint surface\footnote{In the case where $S$ linearly depends on $N$, this condition reduces to $\delta \mathcal{L}(y)/ \delta \phi(x) \neq 0$ on the constraint surface. We note that, though this expression is similar to a secondary constraint $\delta S/ \delta \phi(x)$, $\delta \mathcal{L}(y)/ \delta \phi(x)$ does not need to vanish on the constraint surface when $\mathcal{L}$ includes spatial derivative of $\phi$ as in the case of Horndeski theory.}.
As a result, we do not have any further
secondary (that is, tertiary) constraint and 3 Lagrange multipliers
$v^i$ are undetermined.

Finally, the Poisson brackets among all the constraints can be
summarized in the following table:
	\begin{center}
		\begin{tabular}{c|ccccc|ccc}
			\hline 
			$\left[\cdot,\cdot\right]$ & $\pi_{k}(\vec{y})$ & $p(\vec{y})$ & $p^{kl}(\vec{y})$ & $\pi(\vec{y})$ & $\tilde{\pi}^{kl}(\vec{y})$ & $\mathcal{C}_{k}(\vec{y})$ & $\mathcal{C}(\vec{y})$ & $\frac{\delta S}{\delta\phi(\vec{y})}$\tabularnewline
			\hline 
			$\pi_{i}(\vec{x})$ & 0 & 0 & 0 & 0 & 0 & 0 & 0 & 0\tabularnewline
			$p(\vec{x})$ & 0 & 0 & 0 & 0 & X & 0 & X & X\tabularnewline
			$p^{ij}(\vec{x})$ & 0 & 0 & 0 & 0 & X & 0 & X & X\tabularnewline
			$\pi(\vec{x})$ & 0 & 0 & 0 & 0 & X & 0 & X & X\tabularnewline
			$\tilde{\pi}^{ij}(\vec{x})$ & 0 & X & X & X & X & 0 & X & X\tabularnewline
			\hline 
			$\mathcal{C}_{i}(\vec{x})$ & 0 & 0 & 0 & 0 & 0 & 0 & 0 & 0\tabularnewline
			$\mathcal{C}(\vec{x})$ & 0 & X & X & X & X & 0 & X & X\tabularnewline
			$\frac{\delta S}{\delta\phi(\vec{x})}$ & 0 & X & X & X & X & 0 & X & 0\tabularnewline
		\end{tabular}
		\par\end{center}
where ``X'' stands for generally non-vanishing entries.  It is thus transparent
that in our theory there are $22=17+3+2$ constraints which can be
divided into two classes:
\begin{eqnarray*}
	\text{6 first-class:} &  & \quad\pi_{i},\quad\mathcal{C}_{i},,\\
	\text{16 second-class:} &  & \quad p,\quad p^{ij},\quad\pi,\quad\tilde{\pi}^{ij},\quad\frac{\delta S}{\delta\phi},\quad \mathcal{C}.
\end{eqnarray*}
The number of physical degrees of freedom is calculated to be
\begin{eqnarray}
\#_{\mathrm{dof}} & = & \frac{1}{2}\left(2\times\#_{\mathrm{var}}-2\times\#_{\mathrm{1st}}-\#_{\mathrm{2nd}}\right)\nonumber \\
& = & \frac{1}{2}\left(2\times17-2\times6-16\right)\nonumber \\
& = & 3.
\end{eqnarray}

Thus the degree of freedom due to the higher derivatives, that is the Ostrogradsky's ghost, is absent. We note that if the condition \eqref{det} is not satisfied, additional constraints or gauge degrees of freedom are expected to appear. In such a case, the number of degrees of freedom is less than 3.

\section{Conclusion}
\label{sec:Conclusion}

We have proposed a new class of higher derivative scalar-tensor
theories which break the general covariance. 
The action is given
by (\ref{S_ori}), which can be also viewed as a new type of spatially
covariant gravity theories alternative to those proposed in
\cite{Gao:2014soa,Gao:2014fra,Gao:2018znj}. 
The construction of our
theory is originally motivated from a scalar field with spacelike
gradient while the usual spatially covariant theory in
\cite{Gao:2014soa,Gao:2014fra,Gao:2018znj} is constructed based on a scalar field with timelike gradient. As an illuminating example, we first
gave concrete expressions for the Horndeski action in the spatial gauge.
Motivated with these expressions, we gave a generic action for our new
theory (\ref{S_ori}).  Once the action of the theory is given, it does not rely on that the scalar field gradient is
spacelike. That is, the scalar field in (\ref{S_ori}) can be both spacelike and timelike. The crucial
point is that the scalar field has no kinetic term and hence is non-dynamical. 
Another important point is the spatial diffeomorphism invariance of
the action (\ref{S_ori}). The timelike vector $n^{\mu}$ orthogonal to
some spacelike hypersurface fixes time slice and breaks time
diffeomorphism invariance while keeping the spatial one.  A dynamical scalar degree of freedom from the gravity sector gets excited due to the lack of general covariance. While the presence of the non-dynamical scalar field changes the behaviour of the dynamical scalar mode comparing with that in \cite{Gao:2014soa,Gao:2014fra,Gao:2018znj}. We also make the
Hamiltonian analysis for our action (\ref{S_ori}) and confirm that there are only
three (two tensors and one scalar) dynamical degree of freedom, which ensures the absence of the Ostrogradsky's ghosts.

We notice that a very large class of covariant scalar-tensor theories with higher derivatives fall into (\ref{S_ori}) in the spatial gauge. Our analysis suggests that such theories propagate upto 3 degrees of freedom when we impose the spatial gauge condition, although more degrees of freedom would arise in general. This is simply because, when the spatial gauge is accessible, the kinetic term of the scalar field disappears and the kinetic matrix becomes degenerate automatically, and a single scalar degree of freedom arises due to the breaking of general covariance.

We have a lot of remaining tasks to investigate concerning various aspects of this
new theory, which will be done in the future publications soon. For
example, we are going to make the analysis of cosmological
perturbations. Though the construction was initially inspired by
assuming the presence of a spacelike scalar field, it is
interesting to examine whether this theory still admits cosmological
background solutions, which yield new classes of inflation and/or dark
energy models with new phenomenological features.
Moreover, by construction our theory (\ref{S_ori}) describes a class of scalar-tensor theories wider than DHOST theory in the spatial gauge. It is, however, important to investigate the relationship between our theory and the spatially covariant theory in
\cite{Gao:2014soa,Gao:2014fra,Gao:2018znj}. Another
interesting question is to investigate what happens if we apparently
recover the general covariance of this theory by use of the
St\"{u}ckelberg trick. As for the usual spatially covariant theories, it
was argued that the corresponding generally covariant scalar-tensor theories are healthy at least when the scalar field
gradient is timelike \cite{DeFelice:2018ewo}. Yet another
interesting question is to consider theories with multiple scalar fields
configurations, though, in this paper, we introduce only one
non-dynamical scalar field. All of these topics will be discussed in the
future publications.


\acknowledgments

We would like to thank Rio Saitou and Atsushi Naruko for discussions and
for early collaborations. We also would like to thank Eugeny
Babichev, Christos Charmousis, Marco Crisostomi and Gilles
Esposito-Farese for useful comments and questions.  X.G. was supported
by the Chinese National Youth Thousand Talents Program
(No. 71000-41180003) and by the SYSU start-up funding
(No. 71000-52601106). M.Y. is supported in part by JSPS KAKENHI Grant
Numbers JP25287054, JP15H05888, JP18H04579, JP18K18764, and by the
Mitsubishi Foundation. D.Y. is supported by the JSPS Postdoctoral
Fellowships for Research Abroad.  X.G. would like to thank Masahide
Yamaguchi for hospitality during his stay at Tokyo Institute of
Technology, where the main part of this work was accomplished.

\appendix

\section{$\mathcal{L}_{5}^{\mathrm{H}}$ in the spatial gauge} \label{sec:L5}

In the spatial gauge, the cubic Horndeski Lagrangian (\ref{Horndeski_L5}) reduces to
	\begin{eqnarray}
	\mathcal{L}_{5}^{\text{H,(s.g.)}} & = & \frac{\partial G_{5}}{\partial X}G_{ij}\mathrm{D}^{i}\mathrm{D}^{j}\phi\mathrm{D}^{k}\phi\mathrm{D}_{k}\phi+\frac{1}{2}\frac{\partial G_{5}}{\partial X}\mathrm{D}^{i}\mathrm{D}^{j}\phi\mathrm{D}_{i}\phi\mathrm{D}_{j}\phi\left(K^{2}-K_{kl}K^{kl}\right)\nonumber \\
	&  & -\frac{\partial G_{5}}{\partial X}\left(K_{k}^{\phantom{k}i}K^{kj}\mathrm{D}^{2}\phi-K^{ki}K^{lj}\mathrm{D}_{k}\mathrm{D}_{l}\phi\right)\mathrm{D}_{i}\phi\mathrm{D}_{j}\phi\nonumber \\
	&  & +2\frac{\partial G_{5}}{\partial X}\left(K^{ki}K_{k}^{j}-KK^{ij}\right)\mathrm{D}_{i}\mathrm{D}_{l}\phi\mathrm{D}^{l}\phi\mathrm{D}_{j}\phi\nonumber \\
	&  & -\frac{\partial G_{5}}{\partial X}\left(K^{kl}K^{ij}-KK^{kl}h^{ij}\right)\mathrm{D}_{i}\mathrm{D}_{j}\phi\mathrm{D}_{k}\phi\mathrm{D}_{l}\phi\nonumber \\
	&  & +\frac{1}{3}\frac{\partial G_{5}}{\partial X}\left[\left(\mathrm{D}^{2}\phi\right)^{3}-3\mathrm{D}_{i}\mathrm{D}_{j}\phi\mathrm{D}^{i}\mathrm{D}^{j}\phi\mathrm{D}^{2}\phi+2\mathrm{D}_{i}\mathrm{D}_{j}\phi\mathrm{D}^{j}\mathrm{D}^{k}\phi\mathrm{D}_{k}\mathrm{D}^{i}\phi\right]\nonumber \\
	&  & +\frac{1}{2}\frac{\partial G_{5}}{\partial\phi}\,R\mathrm{D}^{i}\phi\mathrm{D}_{i}\phi-2\frac{\partial G_{5}}{\partial\phi}\,R_{ij}\mathrm{D}^{i}\phi\mathrm{D}^{j}\phi-\frac{1}{2}\frac{\partial G_{5}}{\partial\phi}\left(K^{2}-K_{kl}K^{kl}\right)\mathrm{D}_{i}\phi\mathrm{D}^{i}\phi\nonumber \\
	&  & -2\frac{\partial G_{5}}{\partial\phi}\left(K^{ki}K_{k}^{i}-KK^{ij}\right)\mathrm{D}_{i}\phi\mathrm{D}_{j}\phi+\frac{\partial G_{5}}{\partial\phi}\left[\left(\mathrm{D}^{2}\phi\right)^{2}-\mathrm{D}_{i}\mathrm{D}_{j}\phi\mathrm{D}^{i}\mathrm{D}^{j}\phi\right]\nonumber \\
	&  & +\frac{\partial^{2}G_{5}}{\partial X^{2}}\mathrm{D}_{i}\mathrm{D}_{k}\phi\mathrm{D}^{k}\phi\mathrm{D}^{j}\phi\left(\mathrm{D}^{i}\mathrm{D}_{j}\phi\mathrm{D}^{2}\phi-\mathrm{D}_{l}\mathrm{D}_{j}\phi\mathrm{D}^{i}\mathrm{D}^{l}\phi\right)\nonumber \\
	&  & -\frac{1}{2}\frac{\partial^{2}G_{5}}{\partial X^{2}}\mathrm{D}_{i}\mathrm{D}_{j}\phi\mathrm{D}^{i}\phi\mathrm{D}^{j}\phi\left[\left(\mathrm{D}^{2}\phi\right)^{2}-\mathrm{D}_{k}\mathrm{D}_{l}\phi\mathrm{D}^{k}\mathrm{D}^{l}\phi\right]\nonumber \\
	&  & -2\frac{\partial^{2}G_{5}}{\partial X\partial\phi}\left(\mathrm{D}_{i}\mathrm{D}_{j}\phi\mathrm{D}^{i}\phi\mathrm{D}^{j}\phi\mathrm{D}^{2}\phi-\mathrm{D}_{i}\mathrm{D}_{j}\phi\mathrm{D}^{k}\mathrm{D}^{i}\phi\mathrm{D}^{j}\phi\mathrm{D}_{k}\phi\right)\nonumber \\
	&  & +\frac{1}{2}\frac{\partial^{2}G_{5}}{\partial X\partial\phi}\mathrm{D}^{i}\phi\mathrm{D}_{i}\phi\left(\left(\mathrm{D}^{2}\phi\right)^{2}-\mathrm{D}_{k}\mathrm{D}_{l}\phi\mathrm{D}^{k}\mathrm{D}^{l}\phi\right)\nonumber \\
	&  & +\frac{\partial^{2}G_{5}}{\partial\phi^{2}}\left(\mathrm{D}_{i}\phi\mathrm{D}^{i}\phi\mathrm{D}^{2}\phi-\mathrm{D}_{i}\phi\mathrm{D}_{j}\phi\mathrm{D}^{i}\mathrm{D}^{j}\phi\right). \label{L5H_sg}
	\end{eqnarray}

\section{Higher derivative interactions beyond DHOST theory}
\label{appB}

In this appendix, we show that our general action \eqref{S_ori} includes
the theory which can be obtained by imposing the ``spatial gauge'' from
higher derivative scalar-tensor theories beyond DHOST theory.  Here, as
a simple example, let us consider a scalar-tensor theory which 
non-linearly depends on $\square\phi$,
\begin{align}
 S^{B} = \int \mathrm{d}^4 x \sqrt{-g} F(X, \phi, \Box \phi).\label{SB}
\end{align}
By using \eqref{nabla_2_phi_dec} and imposing spatial gauge condition
\eqref{sg}, we obtain
\begin{align}
 S^{B, (\mathrm{s.g.})} = \int \mathrm{d}^4 x N \sqrt{h}\  F\left(- \frac{1}{2} \mathrm{D}_i \phi \mathrm{D}^{i} \phi, \phi, \mathrm{D}_{i}\mathrm{D}^{i}\phi + \mathrm{D}_{i}\phi  \mathrm{D}^{i} \log{N}\right),\label{SB_sg}
\end{align}
where the acceleration $a_{i}$ is expressed by $N$ through $a_{i} =
\mathrm{D}_{i} \log N$.  Clearly the action \eqref{SB_sg} is included
by our general action \eqref{S_ori}.

\section{Poisson bracket with $\mathcal{C}_{i}$} \label{sec:PBCi}

Let us calculate Poisson bracket between $\mathcal{C}_{i}$ and an
arbitrary function of the phase space variables $Q(\Phi_{I}, \pi^{I})$.
By introducing test functions $\xi^{i}$ and $f$, let us evaluate the
following quantity,
\begin{align}
 \int \mathrm{d}^3 x \mathrm{d}^3 y\, \xi^{i}(x) f(y) [\mathcal{C}_{i}(x), Q(y)]
= [X[\vec{\xi}], {\cal F}],
\end{align}
where ${\cal F}$ is a functional of $f$ and phase space variables $\Phi_{I}$ and $\pi_{I}$, which is given by
\begin{align}
 {\cal F}[f, \Phi_{I}, \pi^{I}] = \int \mathrm{d}^3 y\, f(y) Q(y).\label{calF}
\end{align}
Here we define $f(y)$ so that ${\cal F}$ is invariant under the time
independent spatial diffeomorphism,
\begin{align}
 \delta_{\vec{\xi}} {\cal F} = \int \mathrm{d}^3 x \left(
\frac{\delta {\cal F}}{\delta f} \pounds_{\vec{\xi}} f + \frac{\delta {\cal F}}{\delta \phi_{I}} \pounds_{\vec{\xi}} \phi_{I} + \frac{\delta {\cal F}}{ \delta \pi^{I}} \pounds_{\vec{\xi}} \pi^{I}
\right) = 0.
\end{align}
For example, if $Q$ is a spatial vector $Q^i$, $f$ also has a suffix
and $f_{i}/ \sqrt{h}$ is assumed to transform as a covariant vector.  As
similar way to the derivation of \eqref{PB_LD_xi_com} is given in
\cite{Gao:2018znj}, we obtain
\begin{align}
 [X[\vec{\xi}], {\cal F}] = X[[\vec{\xi},{\cal F}]] + \int \mathrm{d}^3 x \frac{\delta {\cal F}}{\delta f} \pounds_{\vec{\xi}} f.
\end{align}
The first term vanishes because $\vec{\xi}$ is not a phase space
variable. Then, by plugging \eqref{calF} in the second term, we obtain,
\begin{align}
  [X[\vec{\xi}], {\cal F}] = \int \mathrm{d}^3 x\ Q(\vec{x}) \pounds_{\vec{\xi}} f(\vec{x}).\label{XF=Q}
\end{align}
The right hand side vanishes on the constraint surface when $Q(x)$ is a
constraint.  Thus \eqref{XF=Q} guarantees that the Poisson bracket of
$\mathcal{C}_{i}$ with any constraint vanishes on the constraint
surface.

As a consistency check, let us derive the Poisson bracket between
$\mathcal{C}_{i}$ and $N$.
\begin{align}
  \int \mathrm{d}^3 x \mathrm{d}^3 y\, \xi^{i}(\vec{x}) f(\vec{y}) [\mathcal{C}_{i}(\vec{x}), N(\vec{y})]
& = \int \mathrm{d}^3 x N(\vec{x}) \pounds_{\vec{\xi}} f(\vec{x}), \notag\\
&= - \int \mathrm{d}^3 x \pounds_{\vec{\xi}} N(\vec{x})  f(\vec{x}) = - \int \mathrm{d}^3 x \xi^{i}(\vec{x}) f(\vec{x}) \partial_{x^i} N(\vec{x}) \notag\\
&= - \int \mathrm{d}^3 x \mathrm{d}^3 y\, \xi^{i}(\vec{x}) f(\vec{y}) \delta^3(\vec{x}-\vec{y}) \partial_{x^i} N(\vec{x}).
\end{align}
By comparing the expression in the first and last line, we obtain 
\begin{align}
 [\mathcal{C}_{i}(\vec{x}), N(\vec{y})] = - \delta^3(\vec{x}-\vec{y}) \partial_{x^i} N(\vec{x}).
\end{align}
The same expression can be obtained by directly using the definition of $\mathcal{C}_i$, \eqref{calCi_def}. 

\providecommand{\href}[2]{#2}\begingroup\raggedright\endgroup

\end{document}